\pdfoutput=1
\documentclass[sigconf]{acmart}

\usepackage{graphicx}
\usepackage{cases}
\usepackage{multirow}
\usepackage{amsfonts,amssymb}
\usepackage{booktabs}

\usepackage[noend]{algpseudocode}
\usepackage[ruled,linesnumbered]{algorithm2e}

\AtBeginDocument{%
  \providecommand\BibTeX{{%
    \normalfont B\kern-0.5em{\scshape i\kern-0.25em b}\kern-0.8em\TeX}}}

\setcopyright{acmcopyright}
\copyrightyear{2018}
\acmYear{2018}
\acmDOI{10.1145/1122445.1122456}

\acmConference[Woodstock '18]{Woodstock '18: ACM Symposium on Neural
  Gaze Detection}{June 03--05, 2018}{Woodstock, NY}
\acmPrice{15.00}
\acmISBN{978-1-4503-XXXX-X/18/06}

\begin{document}

\title{ISTD-GCN: Iterative Spatial-Temporal Diffusion Graph Convolutional Network for Traffic Speed Forecasting}


\author{Yi Xie}
\affiliation{%
  \institution{School of Computer Science, Fudan University}
  \streetaddress{No.826, Zhangheng Road, Pudong New District}
  \city{Shanghai}
  \country{China}}
\email{18110240043@fudan.edu.cn}

\author{Yun Xiong}
\affiliation{%
  \institution{School of Computer Science, Fudan University}
  \streetaddress{No.826, Zhangheng Road, Pudong New District}
  \city{Shanghai}
  \country{China}}
\email{yunx@fudan.edu.cn}

\author{Yangyong Zhu}
\affiliation{%
  \institution{School of Computer Science, Fudan University}
  \streetaddress{No.826, Zhangheng Road, Pudong New District}
  \city{Shanghai}
  \country{China}}
\email{yyzhu@fudan.edu.cn}







\renewcommand{\shortauthors}{Trovato and Tobin, et al.}

\begin{abstract}
  Most of the existing algorithms for traffic 
  speed forecasting split spatial features and temporal features to independent modules, and then 
  associate information from both dimensions. However, features from spatial and temporal dimensions influence mutually, 
  separated extractions isolate such dependencies, and might lead to inaccurate results. In this paper, we incorporate the 
  perspective of information diffusion to model spatial features and temporal features synchronously. Intuitively, vertices not only diffuse information to 
  the neighborhood but also to the subsequent state along with the temporal dimension. Therefore, we can model such heterogeneous spatial-temporal 
  structures as a homogeneous process of diffusion. On this basis, we propose an \textbf{I}terative \textbf{S}patial-\textbf{T}emporal
  \textbf{Diffusion} \textbf{G}raph \textbf{C}onvolutional \textbf{N}etwork (\textbf{ISTD-GCN}) to extract spatial and temporal features 
  synchronously, thus dependencies between both dimensions can be better modeled. Experiments on two traffic datasets illustrate that our 
  ISTD-GCN outperforms 10 baselines in traffic speed forecasting tasks. The source code is available at https://github.com/Anonymous.
\end{abstract}




\maketitle

\section{Introduction}
Spatial-temporal analysis has received increasing attention with the emergence of big data. 
In the real world, different types of data present dual attributes of spatial and temporal dimensions. 
Thus, spatial-temporal modeling is widely applied in various domains including urban computing \cite{STGCN,DCRNN,graphwavenet}, 
climate science \cite{climatescience,ConvLSTM}, neuroscience \cite{neuapp}, social media \cite{socialapp}, \emph{etc}.
For instance, the traffic speed forecasting task is a typical application of spatial-temporal modeling since 
the traffic speed of specific locations not only depends on the traffic conditions nearby but also on the 
historical traffic information. 

Considerable success has been made in traffic speed forecasting. The mainstream strategy is to incorporate 
convolutional models to handle spatial features (including Convolutional Neural Networks \cite{Grid,gridlike,gridlike2} and 
Graph Convolutional Neural Networks \cite{STGCN,DCRNN,graphwavenet,astgcn}), while sequential models (including Recurrent Neural Networks \cite{DCRNN,structurernn}, 
Temporal Convolutional Neural Networks \cite{STGCN,graphwavenet} and Attention Mechanism \cite{astgcn}) for temporal features.
DCRNN \cite{DCRNN} and Graph WaveNet \cite{graphwavenet} firstly adopt an information diffusion model 
to capture spatial information, which has been proved to be a well simulation of traffic flow in the real world. The perspective of information 
diffusion for spatial features modeling achieves great success.

Nevertheless, a potential pitfall should be noticed: most of the existing models extract 
spatial and temporal information separately, and then associate information from both dimensions by specific methods 
(concatenation, linear transformation, attention mechanism, \emph{etc}.). However, in spatial-temporal data, 
spatial features and temporal features interact mutually, separated extraction might lead to the loss of dependencies, 
and then produce inaccurate modeling.

For removing the weaknesses, we proposed a model called \textbf{ISTD-GCN} (\textbf{I}terative \textbf{S}patial-\textbf{T}emporal
\textbf{D}iffusion \textbf{G}raph \textbf{C}onvolutional \textbf{N}etwork) to capture spatial and temporal features synchronously.
Explicitly, we model traffic flow as a process of information diffusion. Different from DCRNN \cite{DCRNN} and Graph WaveNet \cite{graphwavenet}, 
in our model, the process of information diffusion not only happens in the spatial dimension but also in the temporal dimension simultaneously.
 Vertices diffuse information to the neighborhood but also to the next state along with the temporal dimension. From this perspective, we can 
 capture spatial features and temporal features simultaneously in a unified process of diffusion graph convolution. The pattern 
 of spatial-temporal synchronous diffusion is shown in Figure 1.

The main contributions of our work are as follows:
\begin{itemize}
  \item We incorporate the perspective of information diffusion to model spatial features and temporal features synchronously in spatial-temporal data. 
  Therefore, the dependencies between features from both dimensions have been greatly remained, and more accurate modeling can be generated.
  \item Based on the perspective of spatial-temporal synchronous diffusion, we propose a joint framework \textbf{ISTD-GCN} (\textbf{I}terative \textbf{S}patial-\textbf{T}emporal
  \textbf{D}iffusion \textbf{G}raph \textbf{C}onvolutional \textbf{N}etwork) for modeling traffic speed forecasting task. 
  In ISTD-GCN, we adopts three novel key components for better simulation 
  of spatial-temporal information diffusion and leads to more accurate modeling.
  \item We conduct several experiments with two real-world traffic speed datasets. The experimental results 
  demonstrate that our proposed model outperforms other comparison methods.
\end{itemize}

\begin{figure}[h]
  \centering
  \includegraphics[width=\linewidth]{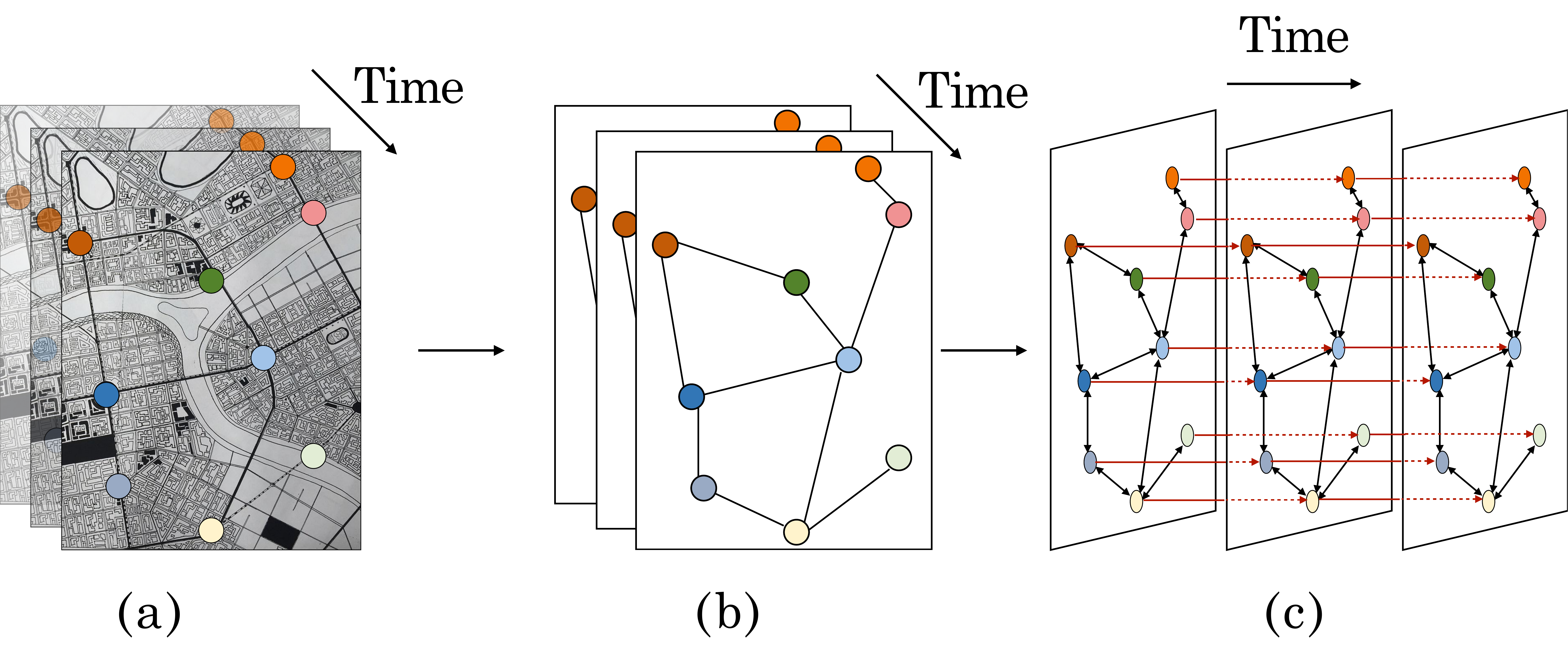}
  \caption{(a) illustrates the raw data collected from the original urban maps. The colored points represent
  points in the city to record traffic speed. The recorded speed will changes with time, while the positions of the colored points are constant.
  (b) shows the original urban map sequence is abstracted as a graph sequence, where the colored points are denoted as vertices in a graph, 
  and the edges encode the pair-wise relationship among colored points in the original urban maps.
  (c) shows the directions of information diffusion in our model. The red lines represent
  the the diffusion along with the temporal dimension, while the black lines represent  
  the diffusion of the spatial dimension. Note that the diffusion along with the temporal dimension is uni-directional.}
\end{figure}

\section{Preliminary}

\subsection{Traffic Speed Forecasting Problem}
The urban road network can be abstracted as a graph, which is represented as $\mathcal{G}=(\mathcal{V},\mathcal{E},W)$,
 where $\mathcal{V}$ is the set of vertices in graph $\mathcal{G}$ with $|\mathcal{V}|=n$, while $\mathcal{E}$ represents
 the set of edges. $W\in{\mathbb{R}^{n\times{n}}}$ denotes the weighted adjacency matrix that is derived from the graph $\mathcal{G}$.
 Entry $W_{ij}$ in matrix $W$ is used to describe the proximity between vertex $i$ and vertex $j$. The detailed construction of 
 the weighted adjacency matrix $W$ is introduced in section 4.1. Note that the topology of graph $\mathcal{G}$ is constant along with the 
 temporal dimension, which means that the weighted adjacency matrix $W$ is also constant. The tensor of dynamic features 
 $X\in{\mathbb{R}^{T\times{n}\times{d}}}$ is the recorded historical observations, where $T$ represents the number of 
 historical time steps, and $d$ denotes the dimension of features. $X^{(t)}$ represents the features of vertices of $\mathcal{G}$ 
 at time step $t$, $\mathcal{G}^{(t)}=(\mathcal{G},X^{(t)})$ is the snapshot of graph $\mathcal{G}$ at time step $t$. 
 The formal definition of traffic speed forecasting problem is to find a mapping function $f(.)$, such that we can infer 
 the snapshots of graph $\mathcal{G}$ in the future $H$ snapshots according to $T$ historical observations:
 \begin{equation}
   f(\mathbb{G}^{(t_{0}:t_{T-1})})=\mathbb{G}^{(t_{T}:t_{T+H-1})},
 \end{equation}
 where $\mathbb{G}^{(t_{0}:t_{T-1})}=(\mathcal{G}^{(t_{0})},\mathcal{G}^{(t_{1})},\cdots,\mathcal{G}^{(t_{T-1})})$ represents the 
 $T$ historical observed snapshots, while $\mathbb{G}^{(t_{T}:t_{T+H-1})}=(\mathcal{G}^{(t_{T})},\mathcal{G}^{(t_{T+1})},\cdots,\\\mathcal{G}^{(t_{T+H-1})})$
 denotes the $H$ future predicted snapshots.

\subsection{Diffusion Graph Convolutional Neural Networks}
Graph neural networks archive great success in handling spatial dependencies in non-Euclidean structures. In our 
model, we refer to the Graph Convolutional Neural Network proposed by \emph{Kipf et al.} \cite{GCN} as the \emph{vanilla GCN}. Graph 
Convolutional Layers in the vanilla GCN are defined as:
\begin{equation}
  X^{(l+1)} = \sigma(\tilde{D}^{-\frac{1}{2}}\tilde{W}\tilde{D}^{-\frac{1}{2}}\Theta^{(l)}X^{(l)}),
\end{equation}
where $X^{(l+1)}$ and $X^{(l)}$ are the output and input for layer $l$, respectively. $\tilde{W}=W+I_{n}$ is the 
weighted adjacency matrix with self-connections, where $W$ denotes the 
weighted adjacency matrix and $I_{n}$ is an identity matrix. $\tilde{D}_{ii}=\sum_j\tilde{W}_{ij}$ is the 
degree matrix of $\tilde{W}$. $\Theta^{(l)}$ is the trainable parameters and $\sigma$ is the activation function.
The item $\tilde{D}^{-\frac{1}{2}}\tilde{W}\tilde{D}^{-\frac{1}{2}}$ denotes the symmetric normalization of the weighted adjacency matrix.

\emph{Li et al.} \cite{DCRNN} proposed a diffusion graph convolutional neural network (\emph{diffusion GCN} for short) for better suiting the nature of 
traffic flow. Specifically, \emph{Teng et al.} \cite{Teng2016ScalableAF} had proved that the diffusion process of information
on graph $\mathcal{G}$ can be formulated as a process of random work with restart probability $\alpha\in{[0,1]}$ 
and transition matrix $(\tilde{D}^{-1}\tilde{W})$. The stationary distribution of 
the diffusion process can be represented
as a weighted combination of infinite random walks on the graph and be calculated in closed form:
\begin{equation}
  \mathcal{P}=\sum^{\infty}_{k=0}\alpha(1-\alpha)^{k}(\tilde{D}^{-1}\tilde{W})^{k},
\end{equation}
where $\mathcal{P}\in{\mathbb{R}^{n\times{n}}}$ denotes the stationary distribution of the Markov process of diffusion.
In DCRNN \cite{DCRNN}, the authors used a finite \emph{K}-step truncation of the diffusion process 
and assigned a trainable weight to each step. Therefore, a step of uni-directional process of information diffusion in DCRNN 
can be rewritten as:
\begin{equation}
  X^{(l+1)} = \sigma(\tilde{D}^{-1}\tilde{W}\Theta^{(l)}X^{(l)}),
\end{equation}

Note that the only difference between Equation (2) and Equation (4) in the form is the approach of normalization, nevertheless,
 there exists fundamental differences on the meanings. In Equation (2), the item $\tilde{D}^{-\frac{1}{2}}\tilde{W}\tilde{D}^{-\frac{1}{2}}$ denotes 
 the symmetric normalization of the weighted adjacency matrix $\tilde{W}$, while the item $\tilde{D}^{-1}\tilde{W}$ in Equation (4)
 denotes the transition matrix in Markov process of information diffusion.

\section{Methodology}

In this section, we formally describe our model: ISTD-GCN. The core idea behind our model is that 
\textbf{\emph{vertices diffuse message not only to the neighborhood but also to the subsequent state along with the temporal dimension.}}
Specifically, there are three key components in our model ISTD-GCN: the Heterogeneous Spatial-Temporal Graph, 
the Two-Step Convolution, and the Iterative Strategy.

\subsection{Heterogeneous Spatial-Temporal Graph}
The Heterogeneous Spatial-Temporal Graph (\emph{HSTG} for short) performs a key role in our model. Spatial-temporal 
data is inherently heterogeneous since features from both dimensions
carry different types of information. Nevertheless, from the perspective of information diffusion, 
the process of diffusion in both dimensions share the homogeneous pattern. The only difference between 
the processes of diffusion on both dimensions is the weights, which control the intensity of signals in the information that 
to be diffused. Therefore, we can connect adjacent snapshots by generating directed edges to its 
subsequent state along with the temporal dimension. The illustration of HSTG that 
is generated by adjacent 3 snapshots is shown in Figure 2.

\begin{figure}[h]
  \centering
  \includegraphics[width=\linewidth]{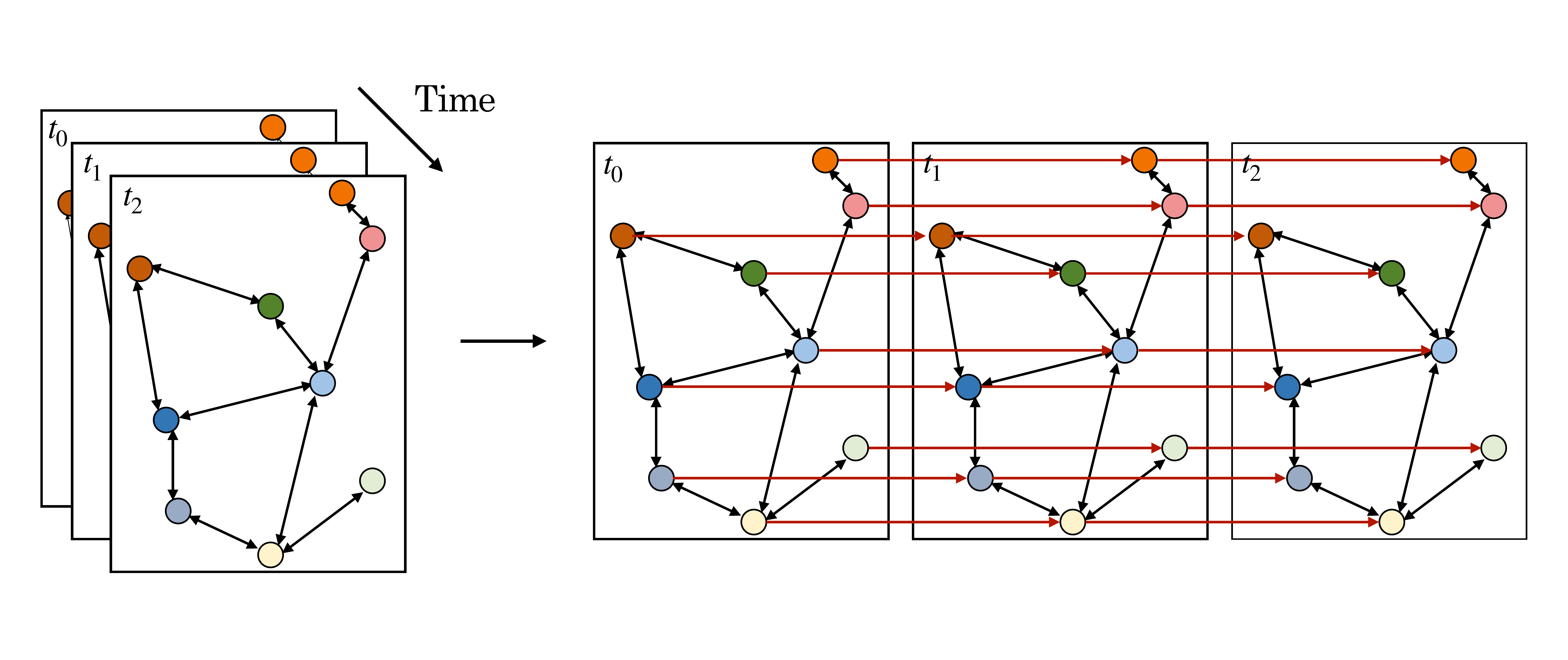}
  \caption{Heterogeneous Spatial-Temporal Graph (HSTG). Black arrows represent bi-directional information diffusion in the spatial dimension, while
  red arrows represent the uni-directional diffusion in the temporal dimension.}
\end{figure}

Formally, we can obtain a sequence of snapshots of graph $\mathcal{G}$ according to $T$ historical observations: 
$\mathbb{G}^{(t_{0}:t_{T-1})}=(\mathcal{G}^{(t_{0})},\mathcal{G}^{(t_{1})},\cdots,\\\mathcal{G}^{(t_{T-1})})$. On 
this basis, we can select arbitrary continuous sub-sequence $\mathbb{G}^{(t':t'+m-1)}=(\mathcal{G}^{(t')},\mathcal{G}^{(t'+1)},\cdots,\mathcal{G}^{(t'+m-1)})$
in $\mathbb{G}^{(t_{0}:t_{T-1})}$ with $m$ snapshots, where $t'\geq t_{0}$ and $t'+m-1\leq t_{T-1}$.
Therefore, we can connect all snapshots in $\mathbb{G}^{(t':t'+m-1)}$ and generate a HSTG, 
the HSTG is denoted as  
$\mathcal{G}_{(H)}=(\mathcal{V}_{(H)},\mathcal{E}_{(H)},W_{(H)})$ with corresponding features $X_{(H)}$, where $\mathcal{V}_{(H)}$ and $\mathcal{E}_{(H)}$ denote vertices and edges in the HSTG, respectively, 
while $W_{(H)}$ is the weighted adjacency matrix of the HSTG. Thus, we can obtain a new graph called Heterogeneous Spatial-Temporal Graph (HSTG) by 
the means of connection of snapshots. The weighted adjacency matrix $W_{(H)}$ of the HSTG is defined as:
\begin{equation}
  W_{(H)}=
  \begin{pmatrix}
    W & C & 0 & \cdots & 0 & 0 & 0 \\
    0 & W & C & \cdots & 0 & 0 & 0 \\
    \vdots &  \vdots & \vdots & \ddots & \vdots & \vdots & \vdots \\
    0 & 0 & 0 &  \cdots & W & C & 0\\ 
    0 & 0 & 0 &  \cdots & 0 & W & C\\ 
    0 & 0 & 0 &  \cdots & 0 & 0 & W\\ 
  \end{pmatrix}
  \in{\mathbb{R}^{mn\times{mn}}},
\end{equation}
where $C$ denotes an matrix consists of the generated edges that connecting all snapshots.

Analogously, the features of corresponding snapshots should also be stacked:
\begin{equation}
  X_{(H)}=STACK(X^{(t')},X^{(t'+1)},\cdots,X^{(t'+m-1)})\in \mathbb{R}^{m\times{n\times{d}}},
\end{equation}
where $STACK(.)$ denotes the operation of stacking.

Intuitively, we can generate edges to connect the corresponding
vertices between different graph snapshots, the generated edges should be directed since time series is uni-directional, which can be 
reflected by the upper triangular adjacency matrix $W_{(H)}$. In addition, The weights of these generated directed edges can be arbitrary since
the intensity of diffused information can be also adjusted by the auto-learned parameters. In our setting, for simplification, 
we set the weights of these generated directed edges as 1. In another word, we specify an identity matrix to connect adjacent snapshots, \emph{i.e.}, 
$C=I_{n}$ in Equation (5), where $I_{n}$ denotes an identity matrix.
Meanwhile, the original undirected graph $\mathcal{G}$ at each snapshots $\mathcal{G}^{(t)}=(\mathcal{G},X^{(t)})$ is also 
transformed into a directed graph, in which the original undirected edges are transformed into bi-directional edges. 
From the weighted adjacency matrix of HSTG $W_{(H)}$ we can also easily obtain the corresponding degree matrix $D_{(H)}$, where $D_{(H)ii}=\sum_{j}W_{(H)ij}$, and 
its inverse matrix $D_{(H)}^{-1}$.

Therefore, the preconditions for performing diffusion GCN described as Equation (4) on HSTG is obtained.

\subsection{Two-Step Convolution}
We can perform diffusion graph convolution on the HSTG $\mathcal{G}_{(H)}$ to simultaneously extract features from the spatial dimension and the temporal dimension 
as following form:

\begin{equation}
  \hat{X}=\sigma{(D_{(H)}^{-1}W_{(H)}\Theta X_{(H)})},
\end{equation}
where $\hat{X}$ denotes the output of the layer.

As we mentioned above, the process of information diffusion in both dimensions is homogeneous. Nevertheless, features in 
both dimensions are not homogeneous since features in different dimensions carry diffenrent 
types of information. In other words, spatial features and temporal features not in the same semantic space. 
In our model, however, is not able to distinguish which semantic space the features come from, since features from both dimensions
share the homogeneous and synchronous process of extraction. This issue might lead to excessive confusion of features from both 
dimensions and cause performance degration.

For handling the problem, we propose a process of two-step diffusion graph convolution (\emph{two-step convolution} for short) to 
force the model distinguish features from different dimensions under the same process of information diffusion.
Specifically, we perform convolution on each snapshot before convolution on HSTG, the purpose is to 
add simple marks for information from different dimensions for identification.
In a nutshell, in each step of our proposed two-step diffusion, we perform information diffusion once in the temporal dimension, 
while twice in the spatial dimension. Therefore, the model can easily distinguish features from both dimensions.
Formally, we disconnect the edges among different snapshots that diffuse the temporal features in HSTG, and generate a 
new graph called non-Heterogeneous Spatial Temporal Graph (\emph{NHSTG} for short) with $m$ connected components, which is denoted as $\mathcal{G}_{(NH)}=(\mathcal{V}_{(NH)},\mathcal{E}_{(NH)},W_{(NH)})$.
The weighted adjacency matrix $W_{(NH)}$ for NHSTG is defined as:
\begin{equation}
  W_{(NH)}=
  \begin{pmatrix}
    W & 0 &  \cdots & 0 & 0  \\
    0 & W &  \cdots & 0 & 0  \\
    \vdots &   \vdots & \ddots & \vdots & \vdots  \\
    0 & 0 &   \cdots & W & 0 \\ 
    0 & 0 &  \cdots & 0 & W \\ 
  \end{pmatrix}
  \in{\mathbb{R}^{mn\times{mn}}},
\end{equation}
analogously, $D_{(NH)ii}=\sum_{j}W_{(NH)ij}$ denotes the degree matrix of $\mathcal{G}_{(NH)}$.
The detailed process of the two-step convolutional layer is shown in Figure 3.
\begin{figure}[h]
  \centering
  \includegraphics[width=\linewidth]{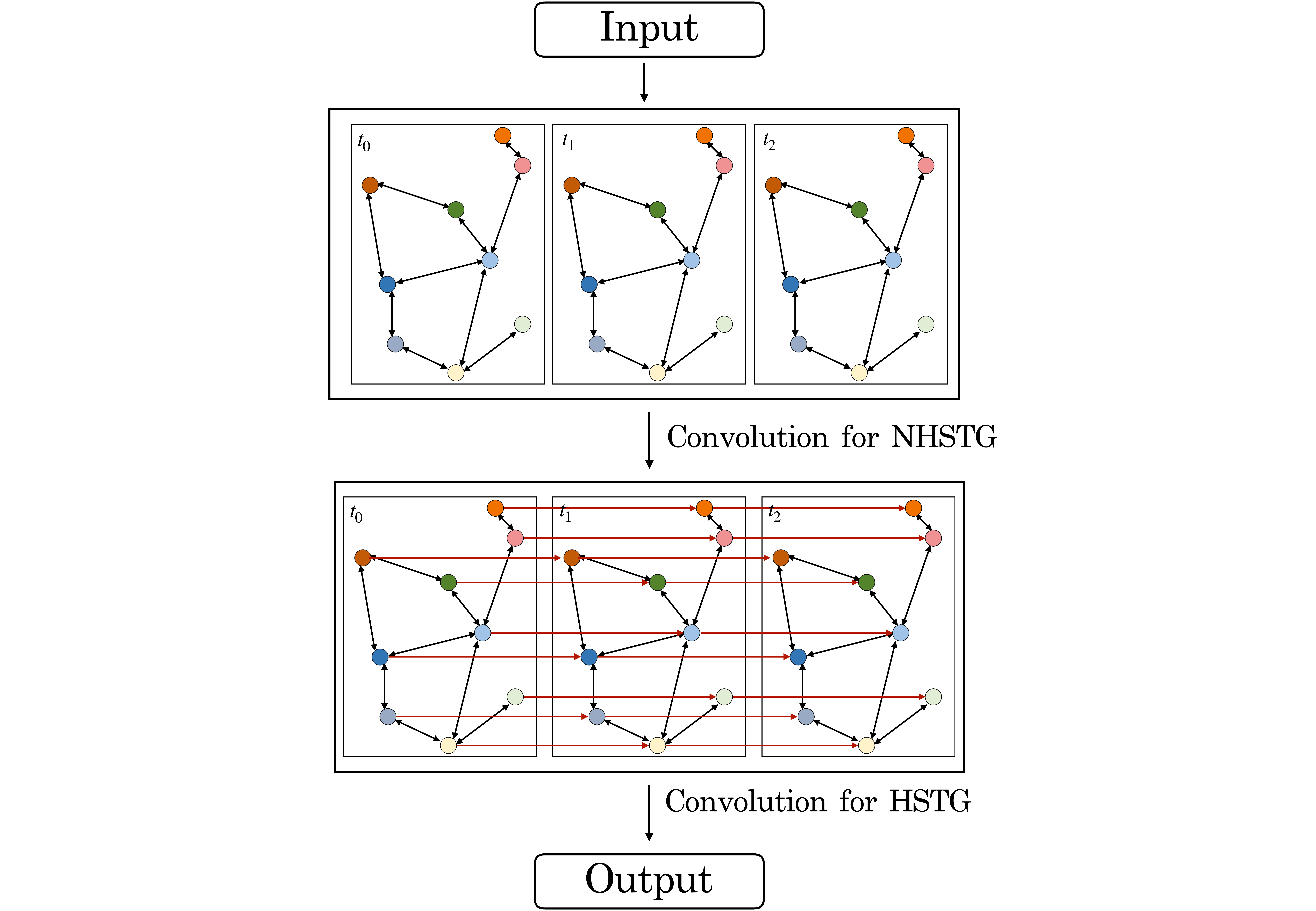}
  \caption{Two-Step Convolutional Layer.}
\end{figure}

We stack several two-step convolutional layers for better performance to generate a spatial-temporal synchronous convolutional block (\emph{STSC block} for short) in our model.
Formally, the naive process of feedforward of our STSC block is defined as:
\begin{equation}
  \hat{X}=\sum_{k=1}^{K}(\sigma((\Theta_{k,1}D_{(NH)}^{-1}W_{(NH)})^{k}X_{(NH)})+\sigma((\Theta_{k,2}D_{(H)}^{-1}W_{(H)})^{k}X_{(H)})),
\end{equation}
where $K$ denotes the number of stacked two-step convolutional layers, \emph{i.e.}, the receptive field of graph convolutional neural networks,
$\Theta_{k,1}$ and $\Theta_{k,2}$ are trainable parameters, while $X_{(NH)}$ and $X_{(H)}$ are input features of convolution on NHSTG and HSTG, respectively.

\begin{figure}[h]
  \centering
  \includegraphics[width=\linewidth]{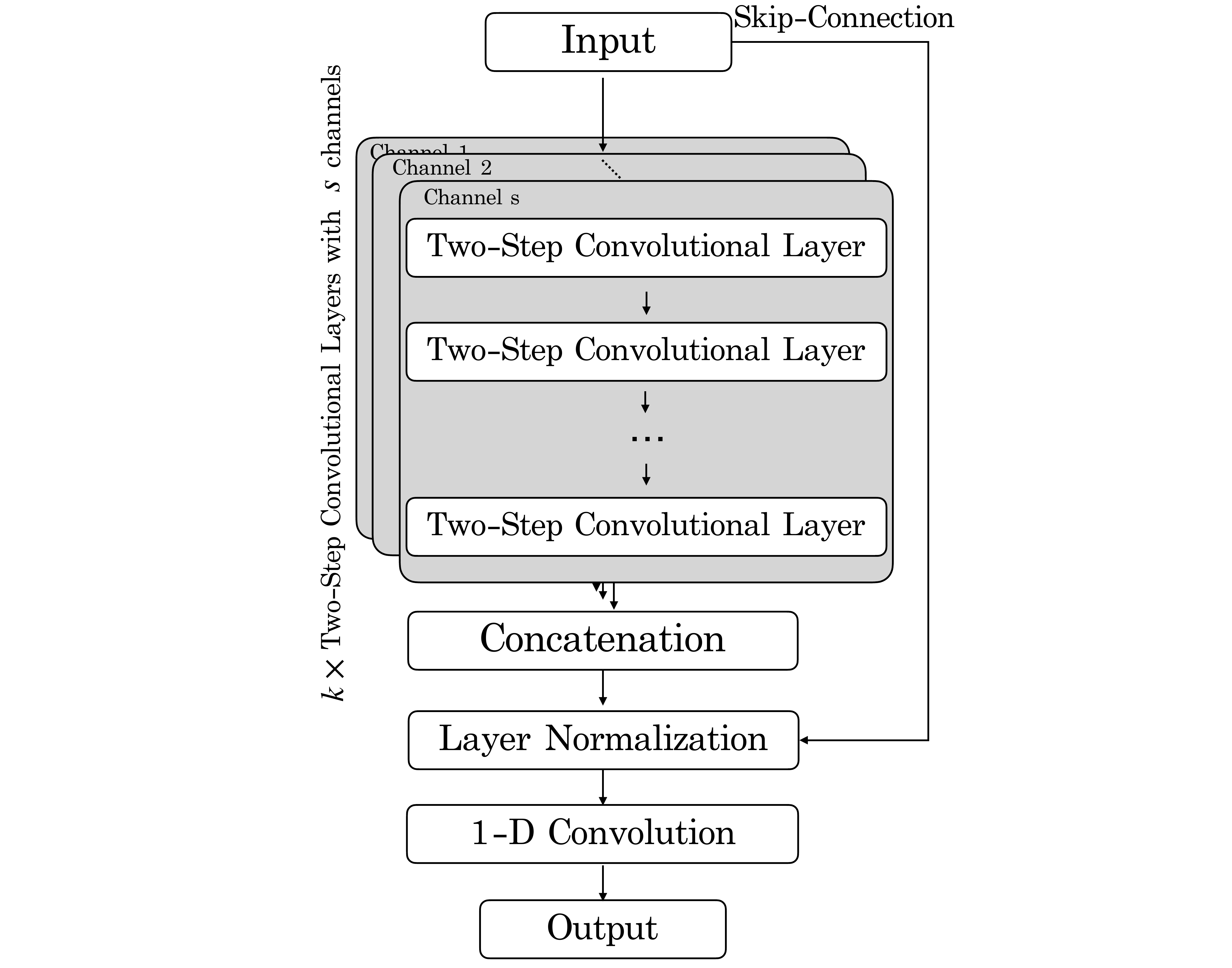}
  \caption{Spatial-Temporal Synchronous Convolutional Block (STSC Block).}
\end{figure}

 The primary purpose of the two-step convolution is to diffuse information instead of non-linear transformation.
Inspired by SGC \cite{SGC}, we removed all non-linear activation layers in the process of convolution for simplification.
For ensuring the stability of representations in hidden layers, we add a layer normalization \cite{layernorm} for the output of 
STSC blocks. We also add skip connection for residual learning to obtain more efficient training \cite{resnet} and alleviate over-smoothing \cite{oversmoothing}.
In addition, the vertices in the NHSTG are the same as the HSTG since we generate the NHSTG by only removing edges. Therefore, $\mathcal{V}_{(NH)}=\mathcal{V}_{(H)}$, corresponding features 
of the NHSTG are also the same as the features of the HSTG, \emph{i.e.}, $X_{(NH)}=X_{(H)}$.
Thus, the propagation rules of our STSC blocks are formally defined as:
\begin{equation}
  \hat{X}=LN(\sum_{k=1}^{K}(\Theta_{k,1}(D_{(NH)}^{-1}W_{(NH)})^{k}+\Theta_{k,2}(D_{(H)}^{-1}W_{(H)})^{k})X_{(H)}+X_{(H)}),
\end{equation}
where $LN(.)$ represents layer normalization. 

Note that the output $\hat{X}\in{\mathbb{R}^{m\times{n}\times{d}}}$ contains features of all vertices in HSTG after STSC block. 
We have to compress the output as a new compressed snapshot since we need to participate
in the subsequent operations. Therefore, we adopt an 1-D convolutional layer to compress the output information. 
Thus, we obtain the compressed feature $\hat{X}\in{\mathbb{R}^{n\times{d}}}$ from the output of STSC block after an 
operation of 1-D convolution. The illustration of a STSC block is shown in Figure 4.

In addition, inspired by the successful application of multi-channel in classical convolutional neural networks \cite{AlexNet},
we adopt several diffenrent graph convolutaion kernels to extract information in different sub-spaces of representations.
The outputs of each channel are concatenated:
\begin{equation}
  \hat{X} = \big|\big|_{i\in{s}}\hat{X}_{i},
\end{equation}
where $\big|\big|$ denotes the operation of concatenation for features, $\hat{X}_{i}\in{\mathbb{R}^{n\times{d}}}$ denotes the output of the $i$-th STSC block, 
$\hat{X}\in{\mathbb{R}^{n\times{sd}}}$ represents the final output of our multi-channel graph convolution, and $s$ is the number of channels.
Therefore, we obtain the output $\hat{X}\in{\mathbb{R}^{n\times{sd}}}$, we adopt a linear transformation to 
reduce the dimension of $\hat{X}$. Finally, the output of STSC block is denoted as $\hat{X}\in{\mathbb{R}^{n\times{d}}}$.

\begin{figure*}[h]
  \centering
  \includegraphics[width=\linewidth]{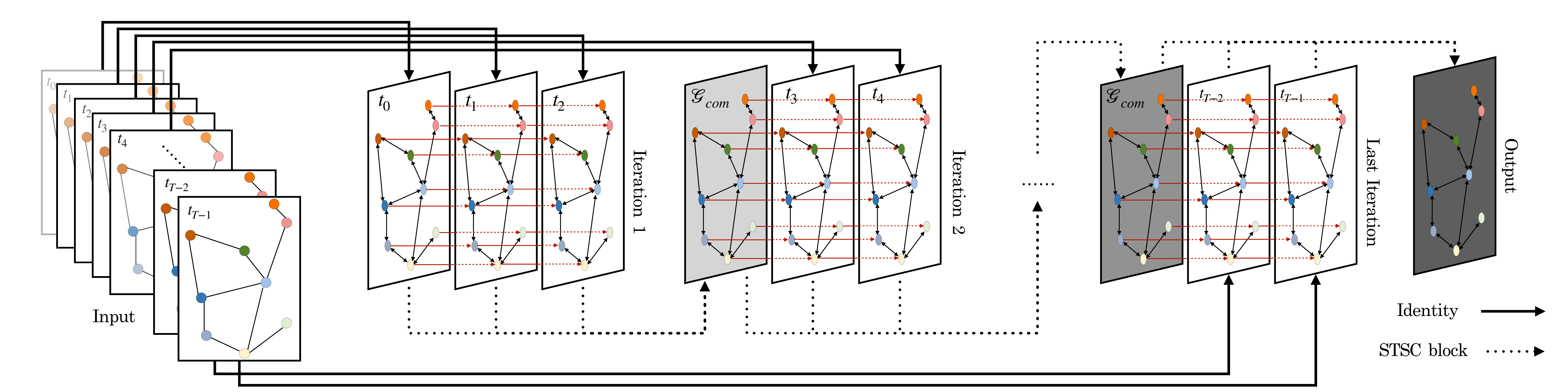}
  \caption{The illustration of Iterative Spatial-Temporal Diffusion Graph Convolution.
  The darker color of graphs represent the compressed snapshots that store and comporess more information in previous snapshots.}
\end{figure*}

\subsection{Iterative Strategy}
Connecting all snapshots to generate larger HSTG and NHSTG in historical observations $\mathbb{G}^{(t_{0}:t{T-1})}$ is not an optimal strategy in 
spatial-temporal data for three reasons:

\noindent1. \textbf{\emph{Overfitting}}. Connecting all snapshots yields a larger trainable parameter matrix as the graph convolutional kernel,
which carries more parameters and leads to an unstable convergence. A larger parameter matrix exacerbates the complexity of our model, and might lead to overfitting.

\noindent2. \textbf{\emph{The loss of local dependencies}}. Intuitively, in the time series, a snapshot is mainly affected by its
adjacent snapshots, and also mainly affects adjacent snapshots, we refer to this property the local dependencies in the temporal dimension. 
In our model, the process of information diffusion 
among adjacent snapshots shares the similar pattern. A lager graph convolutional kernel that caused by 
connecting all snapshots might ignore such dependencies.

\noindent3. \textbf{\emph{The phenomenon of information bottleneck}} \cite{bottleneck}. The structure of our model can be regarded as a variant of Encoder-Decoder architecture \cite{encdec} 
(detailed illustration in section 3.4). Specifically, in the part of the encoder, we encode information in $T$ historical observations into one 
snapshot, and the ratio of compression is $\frac{T-1}{T}$, \emph{i.e.}, the loss of information due to the process of one-time encoding.
As $T$ increases, the loss of information in the process of one-time encoding become unacceptable.

For handling these issues, motivated by previous works that successfully applied 1-D 
convolutional neural networks with smaller convolutional kernels on sequential data \cite{Conv1d1,Conv1d2}, 
we design a strategy of iteration to adapt our model. The detailed process of iterative strategy is shown in Figure 5.

Specifically, in the iteration 1, we select the adjacent $m(m<T)$ snapshots to generate a HSTG (denoted as $\mathcal{G}_{(H)}$) and a NHSTG (denoted as $\mathcal{G}_{(NH)}$), 
then we feed the STSC block on the generated HSTG $\mathcal{G}_{(H)}$ and NHSTG $\mathcal{G}_{(NH)}$ to perform two-step convolution according to Equation (10). 
 Therefore, 
we can obtain a new snapshot $\mathcal{G}_{com}$ that compresses the information from the selected adjacent $m$ snapshots:
\begin{equation}
  \mathcal{G}_{com}=(\mathcal{G},X_{com}^{(t_{0},t_{1},\cdots,t_{m-1})}),
\end{equation}
where $X_{com}^{(t_{0},t_{1},\cdots,t_{m-1})}\in{\mathbb{R}^{1\times{n\times{d}}}}$
 denotes the compressed features that store the information from $X^{(t_{0})}$ to $X^{(t_{m-1})}$.
The topology of $\mathcal{G}_{com}$ equals $\mathcal{G}$.
Then, the compressed snapshot 
$\mathcal{G}_{com}$ is further connected with the following $(m-1)$ snapshots and perform the same operation iteratively. The iteration 
is processing until the end of the sequence of historical observed snapshots $\mathbb{G}^{(t_{0}:t_{T-1})}$.

Thus, we can solve problems caused by connecting all snapshots. Simple illustrations are as following: 

\noindent1. Solution for \emph{overfitting}. The main reason for overfitting is excessive trainable parameters lead to 
extra complexity of model. In the strategy of iteration, the number of entries in graph convolutional kernel is $m^{2}n^{2}$, where $m$ denotes the number of 
connected adjacent snapshots, $n$ is the number of vertices. If we connect all snapshots in the historical observations, the 
number will increases to $T^{2}n^{2}$. Obviously, $m^{2}n^{2} \textless T^{2}n^{2}$ since $m \textless T$. Lower number of 
trainable parameters leads to lower complexity, which can prevent overfitting effectively.

\noindent2. Solution for \emph{the loss of local dependencies}. Abundant different applications of 
Convolutional Neural Networks in images \cite{AlexNet,mobilenet,mobilenetv2} and sequential data \cite{Conv1d1,Conv1d2,graphwavenet,STGCN,wavenet} 
have proved that smaller convolutional kernels can better handle local dependencies, which larger convolutional kernels more 
adapt at extract global features.

\noindent3. Solution for \emph{the phenomenon of information bottleneck}. the phenomenon of information 
bottleneck is caused by excessive information compression in one-time, which might lead to serious loss of information. Specifically, the ratio 
of compression is $\frac{T-1}{T}$ without the strategy of iteration. If we incorporate the strategy of iteration, 
the features will undergo multiple compressions, and the ratio of compression in each of the process is $\frac{m-1}{m}$.
Obviously, $\frac{T-1}{T}>\frac{m-1}{m}$ because $m<T$. Therefore, in each compression, the ratio of 
compression decreases, which means that more useful information is remained and fewer information is lost. Thus, after multiple
processes of compression with lower ratio of compression, we can alleviate the phenomenon of information bottleneck, compared with 
one time with higher ratio of compression.


\subsection{Overall Architecture}
In this subsection, we make a comprehensive description of our model ISTD-GCN. The overall illustration of the model 
is shown in Figure 6. In general, we employ the Encoder-Decoder architecture \cite{encdec}. 

\noindent\textbf{Encoder}: As shown in the Figure 6, the encoder is composed of the part (a), the part (b) and the (c).
Specifically, part (a) is the input sequence of snapshots according to $T$ historical observations: 
$\mathbb{G}^{(t_{0}:t_{T-1})}=(\mathcal{G}^{(t_{0})},\mathcal{G}^{(t_{1})},\cdots ,\mathcal{G}^{(t_{T-1})})$.
The sequence $\mathbb{G}^{(t_{0}:t_{T-1})}$ is fed into the part (b) and perform Iterative Spatial-Temporal Graph Diffusion Convolution, 
which is described detailedly in subsection 3.2 and subsection 3.3. Specifically, the diffusion graph convolution is performed as Equation (10) in each iteration, the iteration
is repeated until the end of the sequence of historical observed snapshots $\mathbb{G}^{(t_{0}:t_{T-1})}$.
The output of the Iterative Spatial-Temporal Graph Diffusion Convolution, and shown in part (c).
The new compressed snapshot that shown in part (c) compresses and stores all information 
of the input sequence of snapshots, which is denoted as $\mathcal{G}_{enc}=(\mathcal{G},X_{enc})$, 
where $X_{enc}\in \mathbb{R}^{1\times{n\times{d}}}$.

\begin{figure}[h]
  \centering
  \includegraphics[width=\linewidth]{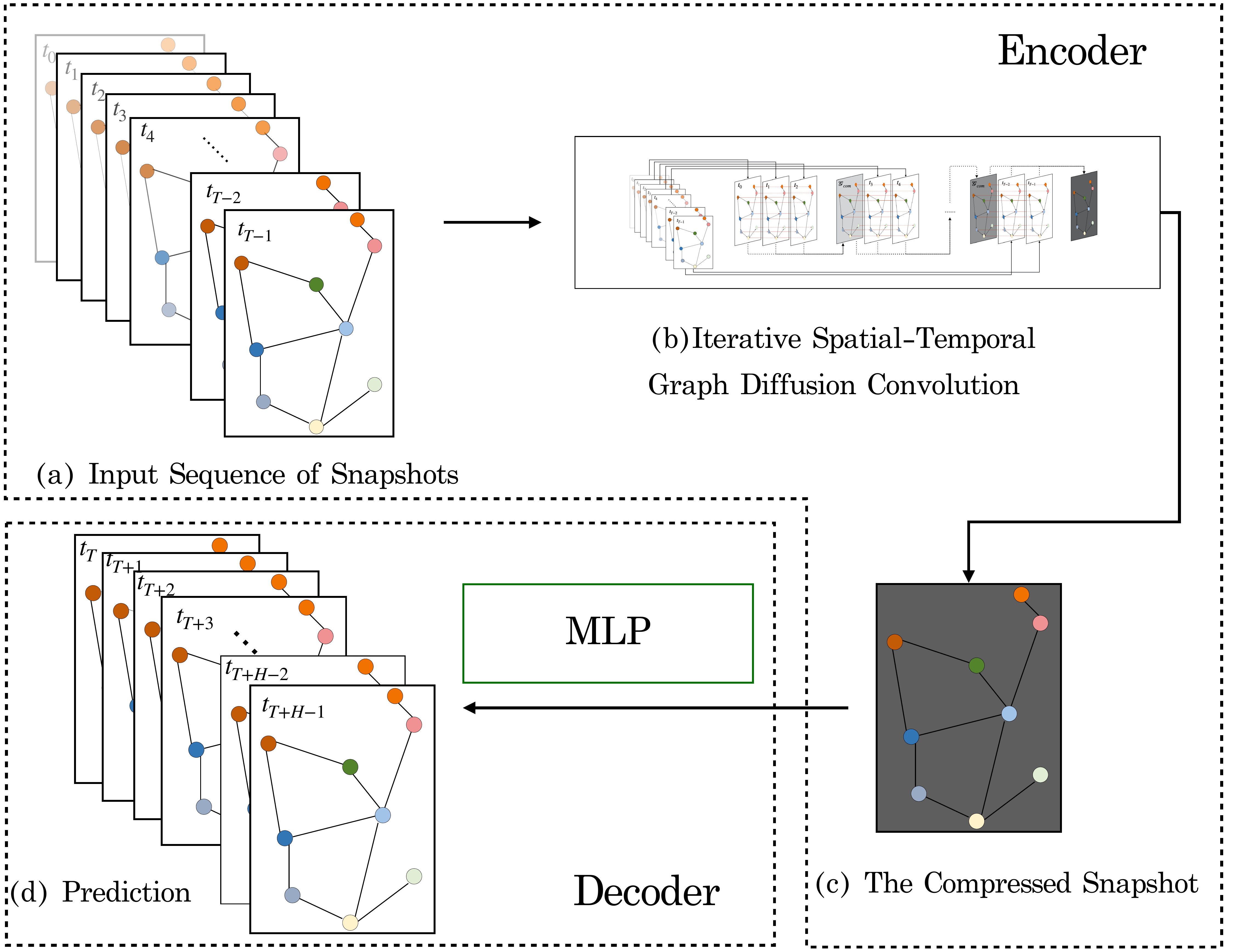}
  \caption{The overview of ISTD-GCN.}
\end{figure}

\noindent\textbf{Decoder}: The decoder is simply but effective. Inspired by \emph{Wu et al.} \cite{graphwavenet},
for eliminating the accumulative error caused by RNN-based decoders, we adopt a Multilayer Perceptron (MLP) to parallelly decode the information 
 at each position. Specifically, the features $X_{enc}\in \mathbb{R}^{1\times{n\times{d}}}$ are transformed into $X_{dec}\in \mathbb{R}^{H\times{n\times{d}}}$ by the MLP.
 Therefore, we can obtain a sequence that composed of 
$H$ snapshots $\mathbb{G}^{(t_{T}:t_{T+H-1})}=(\mathcal{G}^{(t_{T})},\mathcal{G}^{(t_{T+1})},\cdots,\mathcal{G}^{(t_{T+H-1})})=((\mathcal{G},X^{(0)}_{dec}),(\mathcal{G},X^{(1)}_{dec}),\cdots,(\mathcal{G},X^{(H-1)}_{dec}))$ to predict $H$ future conditions, shown as part (d). 

Detailed process of ISTD-GCN is shown as Algorithm 1.

\begin{algorithm}
  \SetAlgoLined
  \KwData{the number of connected snapshots $m$, the sequence of $T$ historical snapshots $\mathbb{G}^{(t_{0}:t_{T-1})}=(\mathcal{G}^{(t_{0})},\mathcal{G}^{(t_{1})},\cdots ,\mathcal{G}^{(t_{T-1})})$.}
  \KwResult{the predicted $H$ snapshots $\mathbb{G}^{(t_{T}:t_{T+H-1})}=(\mathcal{G}^{(t_{T})},\mathcal{G}^{(t_{T+1})},\cdots ,\mathcal{G}^{(t_{T+H-1})})$.}

  \While{not at the end of the sequence $\mathbb{G}^{(t_{0}:t_{T-1})}$}{
    \eIf{is the first iteration}{
      select continuous sub-sequence with $m$ snapshots from the beginning of $\mathbb{G}^{(t_{0}:t_{T-1})}$ as $\mathbb{G}_{s}=\mathbb{G}^{(t_{0}:t_{m-1})}=(\mathcal{G}^{(t_{0})},\mathcal{G}^{(t_{1})},\cdots,\mathcal{G}^{(t_{m-1})})$\; 
      }{
        the current index of iteration is $k$\;
        select continuous  with following $m-1$ snapshots, and concatenate $\mathcal{G}_{com}$ with 
        the selected sub-sequence as $\mathbb{G}_{s}=(\mathcal{G}_{com},\mathcal{G}^{(t_{k(m-1)+2})},\cdots,\mathcal{G}^{(t_{k(m+1)-1})})$\;
      }
      Generate corresponding HSTG $\mathcal{G}_{(H)}=(\mathcal{V}_{(H)},\mathcal{E}_{(H)},W_{(H)})$ and 
      NHSTG $\mathcal{G}_{(NH)}=(\mathcal{V}_{(NH)},\mathcal{E}_{(NH)},W_{(NH)})$ accoding to $\mathbb{G}_{s}$.\;
      Perform the Two-Step Convolution as Equation (10).\;
      Generate the compressed snapshot $\mathcal{G}_{com}$.
      
    }
  $\mathbb{G}^{(t_{T}:t_{T+H-1})}=MLP(\mathcal{G}_{com})=(\mathcal{G}^{(t_{T})},\mathcal{G}^{(t_{T+1})},\cdots ,\mathcal{G}^{(t_{T+H-1})})$ \;
  
  \caption{Iterative diffusion graph convolution}
\end{algorithm}

\subsection{Objective Function}
We incorporate Mean Absolute Error(MAE) as the loss function of ISTD-GCN. Furthermore, for avoiding overfitting, 
we adopt L2 regularization.
\begin{equation}
  L=\frac{1}{HND}\sum_{t=t_{T}}^{t_{T}+H-1}\sum_{n=1}^{N}\sum_{d=1}^{D}|\hat{X}_{nd}^{(t)}-X_{nd}^{(t)}|+\lambda \left\|\Theta\right\|_2,
\end{equation}
where $H$,$N$,$D$ represent the number of predicted snapshots, the number of vertices in $\mathcal{G}$ and the dimension of features, respectively.
$\lambda$ is the coefficient of the L2 regularization and $\Theta$ denotes all trainable parameters in our model.

\section{Experiments}
\subsection{Datasets}
To illustrate the effect of our proposed ISTD-GCN, we process experiments on two real-world traffic datasets METR-LA and PEMS-BAY. The necessary information 
of both datasets is shown in Table 2.

\noindent \textbf{METR-LA} dataset records four months of traffic flow speed information on 207 sensors on the highways of Los Angeles.

\noindent \textbf{PEMS-BAY} dataset records six months of traffic flow speed information on 325 sensors in the Bay area.

For the fairness of the comparison, we adopt the same strategy of data preprocessing with previous works \cite{DCRNN,graphwavenet}.
Specifically,
both datasets aggregate records into 5-minute interval, and 288 snapshots per day. In addition, we adopt the same strategy of using Gaussian kernel \cite{guassiankernel} to construct 
the spatial weighted adjacency matrix:
\begin{equation}
  W_{ij}=
    \begin{cases}
      \exp{(-\frac{d_{ij}^{2}}{\sigma^{2}})}& d_{ij}\leq\epsilon\\
      0& d_{ij}>\epsilon
    \end{cases},
\end{equation}
where $d_{ij}$ denotes the real geographic distance between sensor $i$ and sensor $j$, $\sigma$ is the standard deviation of distances, and 
$\epsilon$ is a threshold. The greater $W_{ij}$ is, the more relevant vertex $i$ and vertex $j$ is. Both settings of $\sigma$ and $\epsilon$ 
are the same as in Graph Wavenet \cite{graphwavenet}.
For the stability of distribution, Z-score normalization
is also applied to the input in order to convert the data distribution to a normal distribution. Z-score is calculated as:
\begin{equation}
  Z=\frac{X-\bar{X}}{S},
\end{equation}
where $X$ denotes the sample to be processed, $\bar{X}$ denotes the mean of samples and $S$ is the standard deviation
of the samples.

\begin{table}
  \caption{Basic information of datasets}
  \label{tab:freq}
  \begin{tabular}{cccc}
    \toprule
    Dataset& \#Vertices& \#Edges& \#snapshots\\
    \midrule
    METR-LA & 207& 1515 & 34272 \\
    PEMS-BAY & 325& 2368 & 52116 \\
  \bottomrule
\end{tabular}
\end{table}

\subsection{Experimental Description and Metrics}
In our experiments, we adopt 12 snapshots (60 minutes) as historical observations and predict traffic speed conditions 
in future 3 snapshots (15 minutes), 6 snapshots (30 minutes) and 12 snapshots (60 minutes), respectively.
We evaluate the performance by three metrics: Mean Absolute Error (MAE), Root Mean Squared Errors (RMSE), and Mean Absolute Percentage Errors (MAPE).
These metrics are calculated as:
\begin{equation}
  MAE(\hat{X},X)=\frac{1}{HND}\sum_{t=t_{T}}^{t_{T}+H-1}\sum_{n=1}^{N}\sum_{d=1}^{D}|\hat{X}_{nd}^{(t)}-X_{nd}^{(t)}|,
\end{equation}
\begin{equation}
  RMSE(\hat{X},X)=\frac{1}{HND}\sum_{t=t_{T}}^{t_{T}+H-1}\sum_{n=1}^{N}\sum_{d=1}^{D}\sqrt{(\hat{X}_{nd}^{(t)}-X_{nd}^{(t)})^2},
\end{equation}
\begin{equation}
  MAPE(\hat{X},X)=\frac{1}{HND}\sum_{t=t_{T}}^{t_{T}+H-1}\sum_{n=1}^{N}\sum_{d=1}^{D}\frac{|\hat{X}_{nd}^{(t)}-X_{nd}^{(t)}|}{X_{nd}^{(t)}+\delta},
\end{equation}
where $\delta$ denotes a tiny shift to prevent the denominator equals zero.

We adopt the same proportion with DCRNN \cite{DCRNN} and Graph WaveNet \cite{graphwavenet} to
 generate training set (60\%), validation set (20\%) and test set (20\%).
We implement our model by PyTorch, and the key hyper-parameters of our model is shown in Table 2.

\begin{table}
  \caption{Hyper-parameters Instruction}
  \label{tab:commands}
  \begin{tabular}{ccc}
    \toprule
    Hyper-parameter & Symbol & Value\\
    \midrule
    The number of layers for two-step convolution & $K$ & 5 \\
    Number of connected adjacent snapshots & $m$ & 2\\
    Number of convolutional channels& $s$ & 8\\
    Dimension of hidden layers & $d$ & 256 \\
    Learning rate & - & 0.0005 \\
    \bottomrule
  \end{tabular}
\end{table}

\subsection{Baselines}
We compared our proposed ISTD-GCN with the following algorithms:\\

\begin{itemize}

  \item \textbf{HA}. Historical Average.
  
  \item \textbf{ARIMA}. Auto-Regressive Integrated Moving Average model with Kalman filter.
  \item \textbf{VAR}. Vector Auto Regression \cite{var}.
  \item \textbf{SVR}. Support Vector Regression which uses linear support vector machine for the regression task. \cite{svr}

  \item \textbf{FNN}. Feedforward Neural Network.
  \item \textbf{FC-LSTM}. Recurrent Neural Network with Fully Connected LSTM hidden units \cite{seq2seq}.

  \item \textbf{WaveNet}. An 1-D convolutional neural network based archietecture for sequence data \cite{wavenet}.
  \item \textbf{DCRNN}. Diffusion Convolutional Recurrent Neural Network. One of the pioneers that incorporating graph neural networks to forecast traffic speed \cite{DCRNN}.
  \item \textbf{STGCN}. Spatial-Temporal Graph Convolution Network. A combination of graph convolutional network and 1-D convolutional network \cite{STGCN}.
  \item \textbf{GraphWavenet}. A combination of graph convolution and dilated casual convolution \cite{graphwavenet}.
\end{itemize}

\subsection{Performance Comparison}
In this subsection, we process detailed performance comparison with aforementioned baselines on dataset METR-LA and PEMS-BAY. 
The results of comparison on both datasets are shown in Table 3 and Table 4, respectively. From the performance comparison, it could be concluded that our proposed ISTD-GCN surpasses other baselines in most of the metrics. Especially,
our algorithm outperforms other competitive algorithms in long-term forecasting tasks, since the parallel decoding 
eliminates the accumulative errors. At the same time, accuracy is maintained since the last compressed output of the encoder compresses all valuable 
information in previous snapshots.

\subsection{Ablation Studies}
We also perform ablation studies to illustrate that three key components in ISTD-GCN is effective. The
 detailed 
results of ablation studies on both datasets are also shown in the of Table 3 and Table 4, respectively.

Specifically, for validating the effect of Spatial-Temporal Synchronous Diffusion, we delete the 
component of Heterogeneous Spatial-Temporal Graph (HSTG) by removing the process of information diffusion in the temporal 
dimension. Analogously, we delete the process of convolution on the non-Heterogeneous Spatial-Temporal Graph (NHSTG) in the 
Two-Step Convolution to illustrate the effect of Two-Step Convolution. Finally, we set the number of connected 
adjacent snapshots $m$ equals the number of historical observations $T$, \emph{i.e.}, $m=T$. Therefore, our algorithm degenerate as a non-iterative algorithm.

\begin{table*}[ht!]
  \caption{Performance comparison and ablation studies on METR-LA}
  \label{tab:datasets}
  \centering
  \begin{tabular*}{\hsize}{@{}@{\extracolsep{\fill}}l|ccc|ccc|ccc}
   \toprule
   \multirow{2}{*}{Model/Condition}&
   \multicolumn{3}{c|}{15 min}&
   \multicolumn{3}{c|}{30 min}&
   \multicolumn{3}{c}{60 min}\\
   \cmidrule{2-10}
   &MAE&RMSE&MAPE(\%)&MAE&RMSE&MAPE(\%)&MAE&RMSE&MAPE(\%)\\
   \midrule
   \midrule
   HA & 4.16 & 7.80 & 13.00 & 4.16 & 7.80 & 13.00 & 4.16 & 7.80 & 13.00 \\
   ARIMA & 3.99 & 8.21 & 9.60 & 5.15 & 10.45 & 12.70 & 6.90 & 13.23 & 17.40 \\
   VAR & 4.42 & 7.89 & 10.20 & 5.41 & 9.13 & 12.70 & 6.51 & 10.11 & 15.80 \\
   SVR & 3.99 & 8.45 & 9.30 & 5.05 & 10.87 & 12.10 & 6.72 & 13.76 & 16.70 \\
   FNN & 3.99 & 7.94 & 9.90 & 4.23 & 8.17 & 12.90 & 4.49 & 8.69 & 14.00 \\
   FC-LSTM & 3.44 & 6.30 & 9.60 & 3.77 & 7.32 & 10.90 & 4.37 & 8.69 & 13.20 \\
   WaveNet & 2.99 & 5.89 & 8.04 & 3.59 & 7.28 & 10.25 & 4.45 & 8.93 & 13.62 \\
   DCRNN & 2.77 & 5.38 & 7.30 & 3.15 & 6.45 & 8.80 & 3.60 & 7.60 & 10.50 \\
   STGCN & 2.88 & 5.74 & 7.62 & 3.47 & 7.24 & 9.57 & 4.59 & 9.40 & 12.70 \\
   GraphWavenet & 2.69 & 5.15 & 6.90 & 3.07 & 6.22 & 8.37 & 3.53 & 7.37 & 10.01 \\
   \midrule
   \textbf{ISTD-GCN} (w/o HSTG) & 2.63 & 4.50 & 6.87 & 2.98 & 5.33 & 8.33 & 3.32 & 5.87 & 9.81 \\
   \textbf{ISTD-GCN} (w/o Two-Step Convolution) & 2.58 & 4.49 & 6.97 & 2.91 & 5.14 & 8.26 & 3.25 & 5.80 & 9.63 \\
   \textbf{ISTD-GCN} (w/o Iterative Strategy) & 2.90 & 5.23 & 8.62 & 3.28 & 5.85 & 9.70 & 3.80 & 6.50 & 11.64 \\
   \midrule
   \textbf{ISTD-GCN}(standard) & \textbf{2.50} & \textbf{4.40} & \textbf{6.50} & \textbf{2.81} & \textbf{5.03} & \textbf{7.37} & \textbf{3.10} & \textbf{5.68} & \textbf{8.91} \\
   \bottomrule
  \end{tabular*}
 \end{table*}

 \begin{table*}[ht!]
  \caption{Performance comparison and ablation studies on PEMS-BAY}
  \label{tab:datasets}
  \centering
  \begin{tabular*}{\hsize}{@{}@{\extracolsep{\fill}}l|ccc|ccc|ccc}
   \toprule
   \multirow{2}{*}{Model/Condition}&
   \multicolumn{3}{c|}{15 min}&
   \multicolumn{3}{c|}{30 min}&
   \multicolumn{3}{c}{60 min}\\
   \cmidrule{2-10}
   &MAE&RMSE&MAPE(\%)&MAE&RMSE&MAPE(\%)&MAE&RMSE&MAPE(\%)\\
   \midrule

   \midrule
   HA & 2.88 & 5.59 & 6.80 & 2.88 & 5.59 & 6.80 & 2.88 & 5.59 & 6.80 \\
   ARIMA & 1.62 & 3.30 & 3.50 & 2.33 & 4.75 & 5.40 & 3.38 & 6.50 & 8.30 \\
   VAR & 1.74 & 3.16 & 3.60 & 2.32 & 4.25 & 5.00 & 2.93 & 5.44 & 6.50 \\
   SVR & 1.85 & 3.59 & 3.80 & 2.48 & 5.18 & 5.50 & 3.28 & 7.08 & 8.00 \\
   FNN & 2.20 & 4.42 & 5.19 & 2.30 & 4.63 & 5.43 & 2.46 & 4.98 & 5.89 \\
   FC-LSTM & 2.05 & 4.19 & 4.80 & 2.20 & 4.55 & 5.20 & 2.37 & 4.96 & 5.70 \\
   WaveNet & 1.39 & 3.01 & 2.91 & 1.83 & 4.21 & 4.16 & 2.35 & 5.43 & 5.87 \\
   DCRNN & 1.38 & 2.95 & 2.90 & 1.74 & 3.97 & 3.90 & 2.07 & 4.74 & 4.90 \\
   STGCN & 1.36 & 2.96 & 2.90 & 1.81 & 4.27 & 4.17 & 2.49 & 5.69 & 5.79 \\
   GraphWavenet & \textbf{1.30} & 2.74 & \textbf{2.73} & 1.63 & 3.70 & 3.67 & 1.95 & 4.52 & 4.63 \\
   \midrule
   \textbf{ISTD-GCN} (w/o HSTG) & 1.53 & 2.66 & 3.83 & 1.82 & 3.20 & 4.13 & 2.11 & 3.82 & 5.18 \\
   \textbf{ISTD-GCN} (w/o Two-Step Convolution) & 1.48 & 2.55 & 3.43 & 1.79 & 3.18 & 4.27 & 2.10 & 3.89 & 5.25 \\
   \textbf{ISTD-GCN} (w/o Iterative Strategy) & 1.66 & 2.80 & 3.78 & 1.86 & 3.34 & 4.33 & 2.07 & 3.98 & 5.17 \\
   \midrule
   \textbf{ISTD-GCN}(standard) & 1.33 & \textbf{2.36} & 2.80 & \textbf{1.61} & \textbf{3.20} & \textbf{3.56} & \textbf{1.90} & \textbf{3.74} & \textbf{4.58} \\
   \bottomrule
  \end{tabular*}
 \end{table*}


 \subsection{Prediction Comparison}

\begin{figure}[h]
  \centering
  \includegraphics[width=\linewidth]{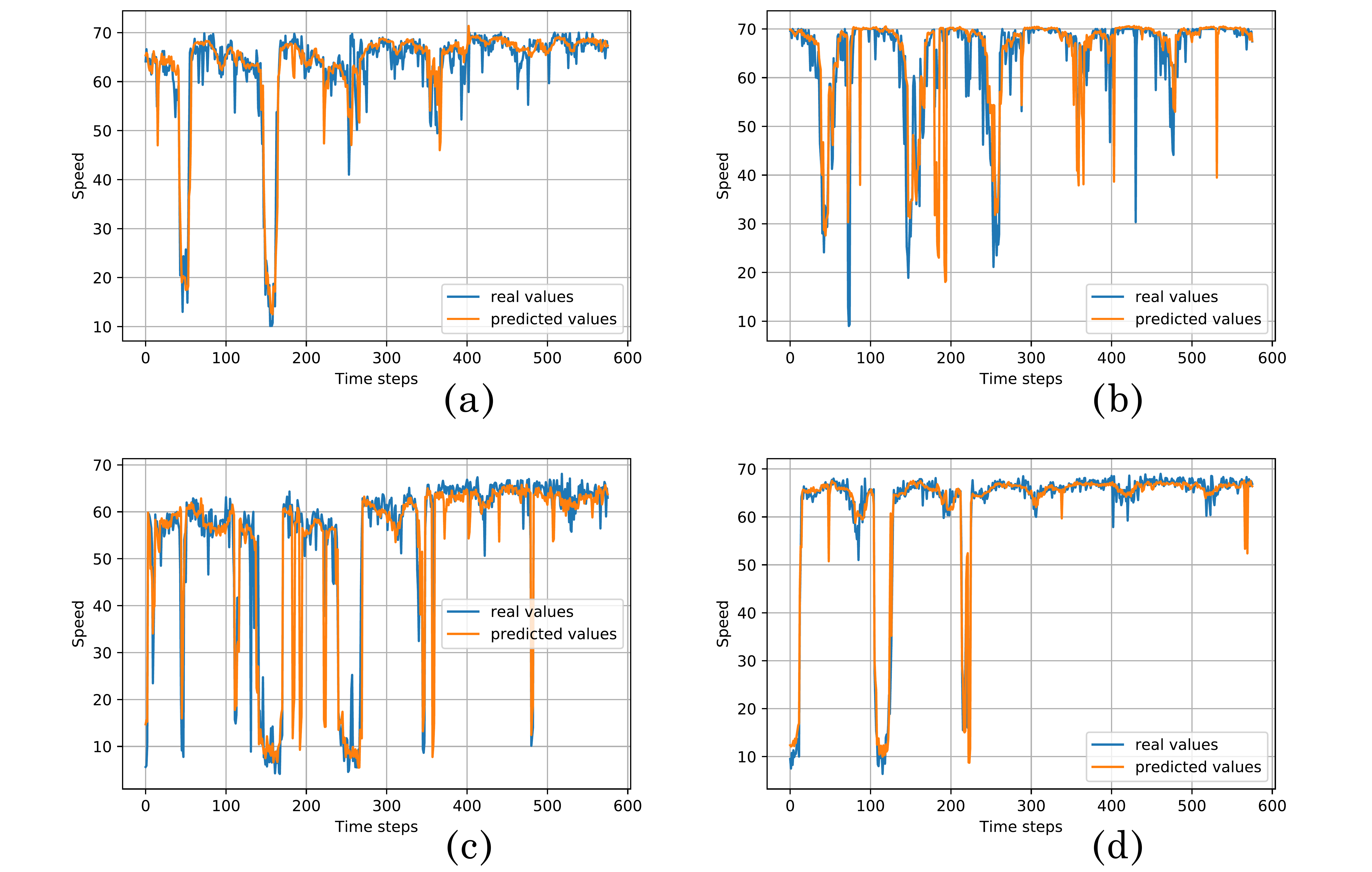}
  \caption{Comparison on 4 randomly selected vertices.}
\end{figure}

\begin{figure}[h]
  \centering
  \includegraphics[width=\linewidth]{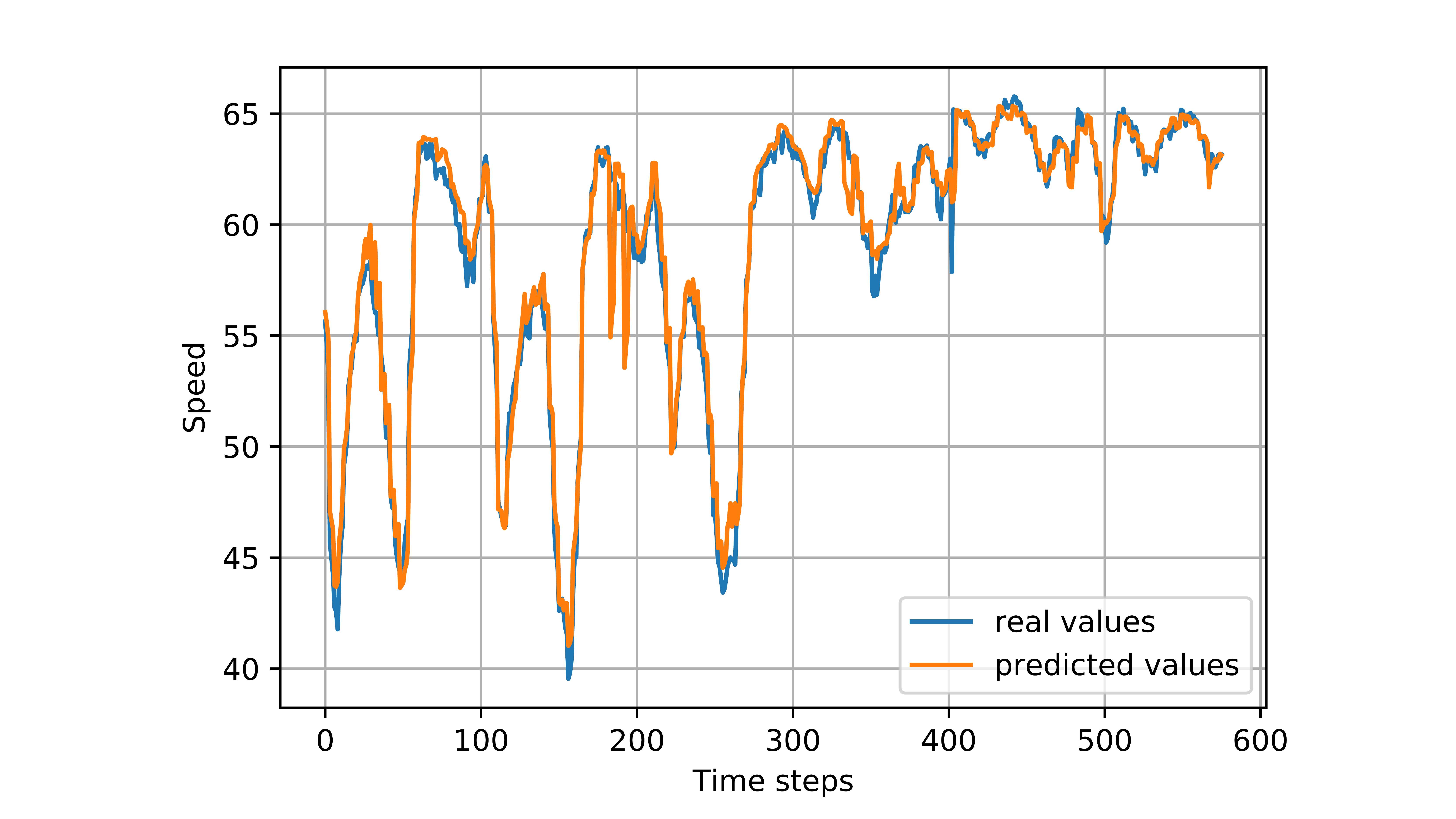}
  \caption{Comparison of average values.}
\end{figure}

 In order to illustrate the performance of our model intuitively, we compared the real values with the predicted values, shown in Figures 7 and 8, respectively.
Specifically, we randomly select 4 vertices and arbitrary 576 continuous snapshots (2 days) to compare the real values and predicted values,
 where the blue lines represent real values while the orange lines denote predicted values. The results of the prediction 
 comparison on specific vertices are shown as Figure 7.
Furthermore, in order to compare the overall prediction on all vertices, we compare the real average speed of all vertices in randomly selected 576 continuous snapshots (2 days)
 with our predicted average speed. The result of prediction comparison of average speed over all vertices is shown as Figure 8.

From Figure 7, we can conclude that our model can fit different patterns of speed change in specific vertices on the micro. From Figure 8, on the macro, our 
model is able to accurately predict future traffic speed over the whole urban road network.

\section{Conclusion}
 In this paper, 
 we incorporate the perspective of information diffusion to model spatial features and temporal features synchronously in spatial-temporal data.
 Therefore, we can better preserve its spatial-temporal dependencies. For handling new issues
 caused by the new perspective, we design three 
 novel key components to improve the performance. 
 On this basis, we propose Iterative Spatial-Temporal Diffusion Graph Convolutional Neural Network (ISTD-GCN). 
 Experiments on two real datasets illustrate that our algorithm outperforms 
   10 baselines in three tasks.

\bibliographystyle{ACM-Reference-Format}
\bibliography{sample-base}










\end{document}